\documentclass[usenatbib]{mn2e}

\usepackage{amsfonts}
\usepackage{amsmath}
\usepackage{amssymb}
\usepackage{graphicx}
\newtheorem{result}{Result}

\title[Radial orbits \& dissipation-induced instabilities]{Radial orbit instability as a dissipation-induced phenomenon}
\author[L.\ Mar\'{e}chal, J.\ Perez]{L.\ Mar\'{e}chal, J.\ Perez\\Laboratoire de Math\'{e}matiques Appliqu\'{e}es, ENSTA, 32 Bd Victor, Paris, France}
\date{Accepted . Received ; in original form }

\begin{document}

\maketitle
\begin{abstract}
This paper is devoted to Radial Orbit Instability in the context of
self-gravitating dynamical systems.
We present this instability in the new frame of Dissipation-Induced Instability
theory. This allows us to obtain a rather simple proof based on energetics
arguments and to clarify the associated physical mechanism.    
\end{abstract}

%---------------------------------------------------------------------

\section{Introduction}
Instabilities in self-gravitating systems are fundamental processes to understand
the shape and physical properties of objects such as galaxies or globular
clusters.
So far, only a few of such mechanisms are described in literature,
namely Jeans instability, which governs the collapse of homogeneous systems;
gravothermal catastrophe, which concerns isothermal spheres;
and radial orbit instability, which occurs in anisotropic, strongly radial
spherical systems.
If the first two are well understood, and 
have taken their place  in the study of dynamical stellar systems
(see \citealt{Binney}, sections 5.2 and 7.3), as a fact, the situation of radial
orbit instability is less clear.
A complete story of this physical process, spanning almost forty years,
is presented in \citet{vlasovia}.
Three main points stick out (see the review for all detailed references):
there is as yet no simple analytical proof of this phenomenon; there is no
global consensus about its actual physical mechanism;
and yet it is a fundamental process which affects the phase space distribution
of primordial galaxies and contributes to produce the radial density profile
of evolved systems.
The present paper will address the first two of these points.

\subsection{Collisionless Boltzmann--Poisson System}
\label{CBPE}

We consider a system constituted of a large number $N$ of gravitating
particles interacting together. We will assume that all those particles have
the same mass $m$. We denote as $\mathbf{q}$ and $\mathbf{p}$ the position and
the associated impulsion of a particle with respect to some Galilean frame
$\mathcal{R}$, and $\mathbf{\Gamma} = \left( \mathbf{q}, \mathbf{p} \right)$
the corresponding point in the phase space $\mathbb{R}^{6}$.

We assume that the statistical state of the system is described at each
instant $t$ by a distribution function $f \left( \mathbf{\Gamma}, t \right)$,
with $f \left( \mathbf{\Gamma}, t \right) \mathrm{d} \mathbf{\Gamma}$
representing the number of particles contained in the elementary phase space
volume $\mathrm{d}\mathbf{\Gamma}$ located around $\mathbf{\Gamma}$.

If the influence of collisions on the overall dynamics is
neglected\footnote{For a self-gravitating system with large values of $N$,
this hypothesis is justified: see \citet{Binney}, part 1.2.1.},
this distribution function solves the Collisionless Boltzmann--Poisson system
(hereafter CBP)
\[
	\left\{ \begin{array}[c]{l}
		\frac{\partial f}{\partial t}
		+ \frac{\mathbf{p}}{m} \cdot \nabla_{\mathbf{q}} f
		- m \nabla_{\mathbf{q}} \psi \cdot \nabla_{\mathbf{p}} f
	= \frac{\partial f}{\partial t} + \left\{ f, E \right\}  = 0\\
	{\displaystyle
		\nabla_{\mathbf{q}}^{2} \psi = 4 \pi G m \int f \mathrm{d} \mathbf{p}
	}
	\end{array}
	\right.
\]
with boundary conditions
$\psi =_{\left\vert \mathbf{q} \right\vert \rightarrow +\infty}
O\left( r^{-1} \right)$
and
$\lim\limits_{\left\vert \mathbf{q} \right\vert ,
\left\vert \mathbf{p}\right\vert \rightarrow +\infty} f = 0$.
The function $\psi \left( \mathbf{q}, t \right)$ is the gravitational potential
created by the particles,
\[
	E\left( \mathbf{q}, \mathbf{p}, t \right) := \frac{\mathbf{p}^{2}}{2m} + m \psi
\]
is the one-particle Hamiltonian, and $\left\{ ., . \right\}$ denotes the Poisson
Bracket defined by
\[
	\left\{ f_{1}, f_{2} \right\}
	= \nabla_{\mathbf{q}} f_{1} \cdot \nabla_{\mathbf{p}} f_{2}
	 - \nabla_{\mathbf{q}} f_{2} \cdot \nabla_{\mathbf{p}} f_{1}
\]

Any stationary solution $f_{0}\left(  \mathbf{\Gamma}\right)$ of the CBP
system is associated to an equilibrium state of the particles distribution. It
is now well known that CBP system is Hamiltonian with respect to a Poisson
bracket of non-canonical form arising from the fact that a distribution function
does not constitute a set of canonical field variables (see \citealt{Kandrup1};
\citealt{Perezaly1}): the set of distribution functions is an
infinite-dimensional space. The total energy
associated to a distribution function $f$ can be written as
\[
	H \left[ f \right]  =
	\int \mathrm{d} \mathbf{\Gamma}
		\frac{\mathbf{p}^{2}}{2m} f \left( \mathbf{\Gamma},t \right)
	- \frac{1}{2} \int \mathrm{d} \mathbf{\Gamma} \int\mathrm{d}\mathbf{\Gamma}^{\prime}
		\frac{f \left(\mathbf{\Gamma},t \right) f\left( \mathbf{\Gamma}^{\prime},t\right)}%
		{\left\vert \mathbf{q} - \mathbf{q}^{\prime} \right\vert }
\]

For any two functionals $A\left[f\right] $ and $B\left[ f\right]$ of the
distribution function, let $\left\langle A,B\right\rangle$ denote the Morrison bracket 
-- introduced in the context of plasma physics by \citet{M80} -- defined by
\[
	\left\langle  A,B \right\rangle =
	\int \mathrm{d}\mathbf{\Gamma} \left\{
		\frac{\delta A}{\delta f}, \frac{\delta B}{\delta f}
	\right\}  f
\]
where $\frac{\delta A}{\delta f}$ stands for the functional derivative of
$A$, which is the linear part of
$A\left[ f+\delta f\right] - A\left[ f\right]$.
One can easily obtain the Hamiltonian formulation of CBP system
\[
	\frac{\mathrm{d}F}{\mathrm{d}t} = \left\langle F,H\right\rangle
\]
where $F\left[ f\right]$ is any functional of $f$.

%---------------------------------------------------------------------

\subsection{The stability problem}

\label{stabproblem}

The stability of equilibrium states is a very old problem of theoretical
stellar dynamics, and a large variety of methods has been used to tackle it.
A clear
consensus was found about the global stability of isotropic spherical systems
with distribution function $f_{0} \left(  E \right)  $ monotonically
decreasing: after the pioneering works by \citet{Antonov}, linear
stability was obtained using energy methods after a long series of papers by
\citet{KS} (and references within) or see also \citet{Perezaly1} for a
comparison of the different results; using direct normal mode techniques,
complicated proofs were also obtained (\citealt{FetP}; \citealt{Palmer}).
Non-linear stability of such spherical isotropic systems was also proven for
some specific models \citep[see][and references within]{Rein}.

The stability of anisotropic spherical systems is a more difficult problem.
The distribution function depends both on the one-particle energy $E$ and on
the squared one particle angular momentum
$L^{2}:=\mathbf{p}^{2}\mathbf{q}^{2} - (\mathbf{p}\cdot\mathbf{q})^{2}$.
The most general result in this context was obtained by \citet{Perezaly1} and
concerns linear stability for the restricted case of preserving perturbations,
which includes radially symmetric ones.
Non-linear
stability is assured for some classes of generalized polytropes for which
$f_{0}\left( E,L^{2}\right) = E^{k}L^{2p}$ with adapted values of $k$ and $p$
(see \citet{Rein} and references within).

Some very technical approaches using normal modes claim linear instability for
anisotropic systems composed only of radial orbits (see \citealt{FetP};
\citealt{Palmer}): this is known as the radial orbit instability.

See \citet{merrit et aguilar} for one of the first
relevant numerical approaches, \citet{perez et al} for an intermediate position
or \citet{ROI Moderne2} and references within for the most recent situation of
this problem.

A complete historical account of radial orbit instability
is given in \citet{vlasovia}.
In the next section, we present some key features of this process.

%---------------------------------------------------------------------

\subsection{Basics of radial orbit instability}
\label{ROI_basis}

Radial orbit instability (hereafter ROI) appears in self-gravitating system dynamics with the pionnering works of \citet{Antonov73} and \citet{henon}. A decade later \citet{polya} propose a stability criterion based on the ratio of radial over tangential kinetic energies, when it is to small the system must leave its spherical symmetry and form a bar. This work is criticized  by \citet{palmer87} which suggest, using normal modes techniques,  that ROI can occur for arbitrary small values of the Russian ratio, provided the distribution function of the system is unbounded for orbits with zero angular momentum. This paper is also the first one to propose a relevant physical mechanism to understand ROI based on resonant trapped orbits; we note that this mechanism needs a coupling between orbits.

Several factors show that a radial system needs a ``seed'' from
which ROI can appear. This is developed in detail in \citet{roy}.
The general idea is that there has to be a near-equilibrium state, so that
coupling between orbits has the time to develop, and the instability to
grow. In \citet{MCMillan}, density profiles in a power law are considered,
which means that the core has the time to stabilize before outer zones
collapse, causing ROI to appear; it also shows that adding clumps
tends to accelerate the process. A counterexample can be found in
\citet{supprROI}, which shows that homogeneous spherical haloes do not
undergo ROI, as the system tends to isotropy before reaching equilibrium. 
For this reason, our study will focus on ROI emerging from equilibrium
states.

Although ROI is a natural candidate to produce triaxiality which can occur in
self-gravitating systems, it was noted (\emph{e.g.} \citealt{Nkatz}) that this
spatial counterpart of ROI could disapear during the merging process of the
galaxy formation. However, \citet{huss} and more recently \citet{MCMillan} have
shown that ROI is a fundamental initial process which shapes the phase space of
the galaxy progenitor and allows it to get the good final mass density profile.
It is therefore important to understand fully the nature of ROI.

In this context, the objective of this paper is twofold. On the first hand, in
section~\ref{section-method}, we present a general method for investigate
instability of self-gravitating systems. This approach couples a general
mathematical result by \citet{BKMR} which generalizes Lyapunov theory, and the
symplectic approach of the stability problem of CBP system (see
\citealt{Bartho}; \citealt{Kandrup1}; and \citealt{Perezaly1}) -- it must be noted
that \citet{Kandrup2} has already used this technique for non spherical
systems without the complete mathematical background.

On the second hand, in section~\ref{section-instab} we apply this method to
obtain a direct energy proof of the radial orbit instability when the system
can dissipate energy.

%---------------------------------------------------------------------

\section{Dissipation-induced instabilities and self gravitating systems}

\label{section-method}

%---------------------------------------------------------------------

\subsection{The method of energy variation}

Consider the first-order variation of an equilibrium $f_{0}\rightarrow
f_{0}+f^{(1)}$. It is well-known (see \citet{Bartho}, \citet{Kandrup1} and
\citet{Perezaly1}) that there exists a phase space function
$g\left(\mathbf{\Gamma},t\right)$, such that
\begin{equation}
	f^{(1)}\left(  \mathbf{\Gamma},t\right) = -\left\{ g,f_{0}\right\}
	\label{symp-pert}
\end{equation}
This function is called a generator\thinspace\footnote{This function is
clearly not unique.} of the perturbation. Written in this form, $f^{(1)}$ is
the largest class of physical perturbations which can be considered as acting
on $f_{0}$. In other words, $f^{(1)}$ is a deformation of $f_{0}$ and then there
exists a $g$ such that we have equation~(\ref{symp-pert}). Associated to this
perturbation, variation of the total energy -- which turns out to be of
second order in $g$, see for instance a short calculation in \citet{vlasovia}
-- is given by
\begin{equation}
	H^{(2)} [f_{0}]
	= -\int \{g, E\} \{g, f_{0}\} \mathrm{d} \mathbf{\Gamma}
	- G m^2 \int\!\!\int \frac{ \{g,f_{0}\} \{g^{\prime},f_{0}^{\prime}\}}%
		{|\mathbf{q} - \mathbf{q}^{\prime}|}
		\mathrm{d}\mathbf{\Gamma}\,\mathrm{d}\mathbf{\Gamma}^{\prime}
\label{h2}
\end{equation}
where $\mathbf{\Gamma}^{\prime}$ refers to
$\left( \mathbf{q}^{\prime}, \mathbf{p}^{\prime} \right)$,
$f_{0}^{\prime}$ to $f_{0}\left( \mathbf{\Gamma}^{\prime} \right)$
and so on.

When $H^{(2)}[f_{0}]$ is positive for a given set $\mathbb{G}$ of acceptable
generators $g$, the system is reputed stable against the associated
perturbations. This argument was detailed and used to prove, in the case
when $f_{0} = f_{0} \left( E \right)$ and
$\partial_{E} f_{0} := \frac{\partial f_{0}}{\partial E} < 0$,
stability against all acceptable $g$; and when
$f_{0} = f_{0} \left(  E, L^{2}\right) $ and
$\partial_{E} f_{0}<0$, stability for all $g$ such that
$\{g,L^{2}\}=0$ which are called preserving perturbations
(see \citealt{Perezaly1} for all details).
\label{cas-dep}

When there are negative energy modes, generators $g$ that cause
$H^{(2)}[f_{0}] < 0$, they are not necessarily associated to an instability.
Taking into account dissipation in the system can drastically change its dynamics.

In the next section, we will illustrate this point with a simpler, yet
instructive example.

%---------------------------------------------------------------------

\subsection{An electromagnetic example of dissipation-induced instability}

\label{section-example}

Consider a particle of mass $m=1$, charge $e$, let us denote as
$\mathbf{q} =  \left( x,y,z \right)^{\top}$ its position with respect to
some Galilean frame.
This particle is influenced by two forces: one derives from a potential $V$
that is maximal at $\mathbf{q}=0$. We'll write
$V(\mathbf{q}) = -\frac{1}{2}\omega^{2}\mathbf{q}^{2}$.
The other one is the Lorentz force generated by a static magnetic field
$\mathbf{B} = B_0 \mathbf{e}_{z} = \mathbf{\nabla}\wedge\mathbf{A}$ with 
$\mathbf{A} = \frac{B_0}{2} \left( x\mathbf{e}_{y} - y \mathbf{e}_{x}\right)$.
The Lagrangian of this particle is
$\mathcal{L}=\frac{1}{2}\dot{\mathbf{q}}^{2} + e\dot{\mathbf{q}} \cdot \mathbf{A}
+ \frac{1}{2}\omega^{2}\mathbf{q}^{2}$,
then the impulsion $\mathbf{p}$ conjugate to the position $\mathbf{q}$ is given by
$\mathbf{p}=\mathbf{\nabla}_{\dot{\mathbf{q}}}\left( \mathcal{L}\right)
= \left(p_{x},p_{y},p_{z}\right)^{\top}$, and thus, with
$\beta = \frac{e B_0}{2}$:
\[
	\left\{ \begin{array}[c]{l}
		p_{x}=\dot{x}-\beta y\\
		p_{y}=\dot{y}+\beta x\\
		p_{z}=\dot{z}
	\end{array} \right.
\]
The Hamiltonian of the system is
$\mathcal{H}=\mathbf{p} \cdot \dot{\mathbf{q}} - \mathcal{L}$,
the equations of motion are given by Hamilton's ones, i.e.
$\dot{\mathbf{q}} = \mathbf{\nabla}_{\mathbf{p}}\left( \mathcal{H}\right)$
and $\mathbf{\dot{p}}= -\mathbf{\nabla}_{\mathbf{q}}\left( \mathcal{H}\right)$.
The behaviour of $\left( z,p_{z}\right)$ being trivial and independent from
movement on the other axes, let us focus on the system
in the reduced phase space of $\xi=\left( x,y,p_{x},p_{y}\right)^{\top}$ for
which one has
\begin{eqnarray}
	\dot{\xi}=\Lambda\xi & \text{where} &
	\Lambda = \left( \begin{array}[c]{cc}
		\beta K & I_{2}\\
		\alpha I_{2} & \beta K
	\end{array} \right)
	\label{eqmatrix}
	\\
	& \text{with} & K = \left( \begin{array}[c]{cc}
		0 & 1\\
		-1 & 0
	\end{array} \right),
	\quad \alpha = \omega^{2}-\beta^{2}
\end{eqnarray}

Since $K^2=-I_2$ one can see that if we split $\Lambda=A+B$ with
\[
	A=\left( \begin{array}[c]{cc}
		0 & I_{2}\\
		\alpha I_{2} & 0
	\end{array} \right)
	\quad \text{and} \quad
	B = \beta\left( \begin{array}[c]{cc}
		K & 0\\
		0 & K
	\end{array} \right)
\]
we have the fundamental property $AB = BA$, so 
$\exp\left(\Lambda t\right)  =\exp\left( At\right)  \exp\left( Bt\right)$,
by direct series summation one can find that
\[
	\exp\left(  Bt\right) =
	\left( \begin{array}[c]{cc}
		\Sigma\left( t\right)  & 0 \\
		0 & \Sigma\left( t\right)
	\end{array} \right)
\]
with
\[
	\Sigma\left( t\right) = \left( \begin{array}[c]{cc}
		\cos\left(  \beta t\right)  & \sin\left(  \beta t\right) \\
		-\sin\left(  \beta t\right)  & \cos\left(  \beta t\right)
	\end{array} \right)
	\in SO_{2}\left( \mathbb{R}\right)
\]
and
\[
	\exp\left(  At\right) =
	\left( \begin{array}[c]{cc}
		\varphi_{1}\left(  t\right)  I_{2} & \varphi_{2}\left(  t\right)  I_{2}\\
		\alpha\varphi_{2}\left(  t\right)  I_{2} & \varphi_{1}\left(  t\right)  I_{2}
	\end{array} \right)
\]
with
\begin{equation*}
	\begin{array}{lcl}
		\left\{ \begin{array}{ccl}
			\varphi_{1}\left(  t\right) & = &
			\cosh\left(  \sqrt{\alpha} t\right)\\
			\varphi_{2}\left(  t\right) & = &
			( \alpha)^{-1/2}\sinh\left(  \sqrt{\alpha} t\right)
		\end{array} \right.
		& \text{if} & \quad \alpha > 0\\
		& &
		\\
		\left\{ \begin{array}{ccl}
			\varphi_{1}\left(  t\right) & = &
			 \cos \left(  \sqrt{-\alpha}t\right)\\
			\varphi_{2}\left(  t\right) & = &
			(-\alpha)^{-1/2}\sin \left(  \sqrt{-\alpha}t\right)
		\end{array} \right.
		& \text{if} & \quad \alpha\leq 0
	\end{array}
\end{equation*}

The general solution of the problem then writes
\[
	\xi\left(  t\right)
	= \exp\left( At\right) \cdot \exp\left(  Bt\right)
		\cdot \xi\left( t=0\right)
\]
and it is stable provided that $\alpha=\omega^{2}-\beta^{2} \leq 0$. 
We note that this stability is not asymptotic as all eigenvalues of $\Lambda$ lie
on the imaginary axis. The mathematical condition on $\alpha$ corresponds to
the physical case when the effect of the magnetic field $\mathbf{B}$ is
stronger than the effect of the scalar potential $V$.
We can see on this example that it is
possible to have a stable equilibrium even on a point where the potential is
at a maximum; \emph{negative energy variations around an equilibrium is not a
sufficient criterion for an instability}. The physical explanation is that
the magnetic force, which does not derive from a scalar potential, tends to
`curve' the particle's trajectory, and if the magnetic field is strong enough,
this can be enough to keep the particle close to the potential maximum in spite
of the repulsive force.

However, this behaviour is only possible as long as there is no energy
dissipation. If the system is able to dissipate energy, such an equilibrium
becomes unstable. For example, assume
there is some form of fluid friction force
$\mathbf{F}_{f}=- \gamma \dot{\mathbf{q}}$, the system is no longer Hamiltonian.
Movement is still trivial in the $\left( z,p_{z}\right)$ plane;
keeping the same variables that we have used in the non-dissipative case,
the equation of motion is now
\[
	\dot{\xi}=\Lambda_{\gamma} \xi
	\quad \text{ where} \quad
	\Lambda_{\gamma} = \Lambda + \gamma C ,
	\quad
	C = \left( \begin{array}[c]{cc}
		0 & 0\\
		-\beta K & -I_{2}
	\end{array}\right) \text{.}
\]
The matrix $C$ happens to commute with $B$, so we have
$(A+\gamma C)B = B(A+\gamma C)$: the fundamental matrix of the dissipative
system splits into
\[
	\exp\left(  \Lambda_{\gamma} t \right)
	= \exp\left( \left[ A+\gamma C\right] t\right)
		\cdot \exp\left(  Bt\right)
\]
As $\exp\left(  Bt\right)  $ is a rotation matrix, the stability of the
dynamics is governed by $\exp\left(  \left[  A+\gamma C\right]  t\right)$.
The characteristic polynomial of $A_{\gamma}=A+\gamma C\ $ is
\[
	\chi(\lambda)
	= \lambda^{4} + 2\gamma\lambda^{3}
		+ (\gamma^{2}-2\alpha) \lambda^{2}
		- 2 \alpha\gamma\lambda
		+ \gamma^{2}\beta^{2}+\alpha^{2}
\]
roots of which are
\[
	\lambda_{1,2} = \frac{1}{2} \left[ -\gamma
		\pm \sqrt{\gamma^{2}+4\alpha + 4i\beta\gamma}\right]
\]
and
\[
	\lambda_{3,4} = \frac{1}{2} \left[ -\gamma
		\pm \sqrt{\gamma^{2}+4\alpha - 4i\beta\gamma}\right]  \text{.}
\]
Let us focus on the transition from the stable equilibrium we have determined
towards the dissipative case. We then have $\alpha=\omega^{2}-\beta^{2}<0$
and $0<\gamma\ll1$. In this limit case, one can get
\[
	\lambda_{1,2} = \frac{\gamma}{2} \left(-1
		\pm \left( 1 - \frac{\omega^2}{\beta^2} \right)^{-1/2} \right)
	\pm i\left( \beta^{2}-\omega^{2}\right)^{1/2}
	+ o\left(\gamma\right)
\]
and
\[
	\lambda_{3,4} = - \frac{\gamma}{2} \left(-1
		\pm \left( 1 - \frac{\omega^2}{\beta^2} \right)^{-1/2} \right)
	\pm i\left( \beta^{2}-\omega^{2}\right)^{1/2}
	+ o\left(\gamma\right)
\]
From our assumption that $\alpha < 0$, we have
% $\left( 1 - \frac{\omega^2}{\beta^2} \right)^{-1/2} > 1$, therefore
$( 1 - \frac{\omega^2}{\beta^2} )^{-1/2} > 1$, therefore
there is a pair of roots \textbf{(eigenvalues of $A_\gamma$)}
with positive real parts.
\textbf{From the equation of motion it follows that the system is unstable;
this kind of instability, linked to its operator's spectrum, is called a
\emph{spectral instability}.}
When $\gamma$ is not infinitesimal, this instability
persists as one can check by direct spectrum calculation or by more elegant
approaches. The physical meaning is clear: if the particle loses energy,
the magnetic field cannot `curve' it back as close to the maximum
as it was previously, and it will spiral further and further from the origin.

%---------------------------------------------------------------------

\subsection{Dissipation-induced instabilities}

\bigskip The previous three-dimensional example is a special case of a general
theorem  which applies for finite dimensional systems:
a Hamiltonian dynamical system with a negative energy mode
(which could be stable without further hypothesis)
becomes spectrally and hence linearly and non-linearly unstable when any
kind of dissipation is introduced.
This counterintuitive result takes its genesis from the classical
works by Thomson (Lord Kelvin) and Tait (1879), but it was proven only
recently in the case of finite-dimensional systems (\citealt{BKMR} and
\citealt{KM07}), and, as suggested by references in the latter,
appears to be very useful in mechanics.
More recent works by \citet{KM09} suggest that the infinite-dimensional
case works similarly, although there is no definitive proof
for the time being.
In the context of theoretical astrophysics, it is interesting to
note that H.~Kandrup used such kind of arguments to investigate gravitational
instabilities for triaxial systems \citep[see][]{Kandrup2}, before any
actual, formal result.

As recalled in section \ref{CBPE}, CBP \emph{is} a Hamiltonian
infinite-dimensional system, so we can apply this theory of
dissipation-induced instability for stability investigations in this
context of gravitational plasmas. In the next section we will show that, when a
spherical and anisotropic self-gravitating system becomes more and more radial,
we can choose a certain class of $g$ for which $H^{(2)}[f_{0}]<0$: this proves
the existence of negative energy modes in such systems.
Following the dissipation-induced
instability theory such kind of gravitating systems will become unstable as
soon as any kind of dissipation can appear. As noticed by Kandrup in his
visionary paper, in physical self-gravitating systems dissipation could take
several forms like a little bit of gas, dynamical friction or at minimum
gravitational radiation! In the context of numerical modelizations of
self-gravitating systems where radial orbit instability also appears,
dissipation is also inevitably introduced by numerical algorithms of
time integration or by potential computation.

%---------------------------------------------------------------------

\section{Application to radial orbit instability}
\label{section-instab}

\subsection{Approaching a radial system}

A pure radial orbit system is characterized by particles with $L^{2}=0$, the
corresponding distribution function could then be written
$f_{0}^{\textbf{ro}} \left(E, L^{2} \right)
= \varphi \left(E \right) \delta\left(L^{2}\right)$
where $\varphi$ is any positive smooth normalized function, and $\delta$ denotes the
Dirac distribution. However, this distribution is very irregular in zero which
is quite problematic, in addition to being unrealistic (orbits can hardly be
perfectly radial). So, instead of actually using the Dirac distribution, we
will use functions that approach it.

The choice we made is to use Gaussian functions.
More specifically, we will consider an initial distribution function of the form
\begin{equation}
	f_{0}^{a} \left( E, L^{2}\right)
	= \varphi\left( E\right) \delta_{a}\left(L^{2}\right),
	\quad 
	\delta_{a}(L^{2}) = \frac{1}{\pi a^{2}} \exp\left(-\frac{L^{2}}{a^{2}}\right)
\label{radial-form}
\end{equation}

By direct calculation one can easily check that, for any smooth function $Z$
defined on the phase space, one has, by limited development of $Z$ with
respect to $p_\theta$ and $p_\phi$%
\footnote{N.B.: variables $p_r$, $p_\theta$ and $p_\phi$ thereafter are the
conjugate variables of $r$, $\theta$ and $\phi$, not the projections of
$\mathbf{p}$ along the base vectors. We have
$p_r = m \dot{r}$, $p_\theta = m r^2 \dot{\theta}$ and
$p_\phi = m r^2 \sin^2(\theta) \dot{\phi}$.
Also
\[
	L^2 = p_\theta^2 + \frac{p_\phi^2}{\sin^2(\theta)}
\]
}$^{,\;}$\footnote{The calculation involves the well-known Gaussian integrals
\begin{eqnarray*}
	\int e^{-\frac{x^2}{r^2}} \mathrm{d}x 
		& = & r \sqrt{\pi} \\
	\int x^2 e^{-\frac{x^2}{r^2}} \mathrm{d}x 
		& = & \frac{1}{2} r^3 \sqrt{\pi} \\
	\int x^4 e^{-\frac{x^2}{r^2}} \mathrm{d}x 
		& = & \frac{3}{4} r^5 \sqrt{\pi}
\end{eqnarray*}}
\begin{eqnarray}
	& & \int Z \delta_{a}(L^{2}) \mathrm{d} \Gamma
	 =  \int Z \delta_{a}(L^{2})
		\frac{\mathrm{d}p_{r}\, \mathrm{d}p_{\theta}\,\mathrm{d}p_{\phi}}%
			{r^{2}\sin\left(\theta\right)} \,\mathrm{d}^3\mathbf{q} \nonumber\\
	& = & \int \frac{1}{r^2} \left. \left(
		Z
		+ \frac{a^2}{4} \frac{\partial^2 Z}{\partial p_\theta^2}
		+ \frac{a^2}{4} \sin^2(\theta) \frac{\partial^2 Z}{\partial p_\phi^2}
	\right) \right|_{L^2 = 0}
	\mathrm{d}p_{r}\,\mathrm{d}^3\mathbf{q}
	\nonumber \\
	& & + O(a^4)
\label{intdelta}
\end{eqnarray}
which has a clear limit when $a \rightarrow 0$ and selects the value of
$Z$ at $L^{2}=0$ as expected. We could say that $f_0^a$ tends to a distribution
function of purely radial orbits when $a\rightarrow 0$.

In the following part, we will consider what happens for arbitrarily small
values of $a$.

%---------------------------------------------------------------------

\subsection{Energy variation}

Our goal in this section is to show that there exist perturbation generators
$g$ that, for sufficiently small values of $a$ (that is, for systems that are
close enough to the purely radial case), give a negative energy variation.
To do so, we will start with a general $g$, calculate the energy variation
$H^{(2)}[f_0^a]$ for our quasi-radial systems, and explain along
the way what hypotheses we make about $g$ to reach this goal.

Usual Poisson bracket properties give, for $f_0^a(E,L^2)$
\[
	\{g,f_0^a\} = \partial_E f_0^a \{g,E\} + \partial_{L^{2}} f_0^a \{g,L^{2}\}
\]
where $\partial_E f_0^a := \frac{\partial f_0^a}{\partial E}$ and
$\partial_{L^{2}} f_0^a := \frac{\partial f_0^a}{\partial L^{2}}$,
hence the second order energy variation~(\ref{h2}) splits into
\begin{equation}
	H^{(2)}[f_0^a]
	= K_{L^{2}} + K_{E}
	- G \int\!\!\int\ \frac{(\delta\rho_{L^2}+\delta\rho_{E})%
		(\delta\rho_{L^2}^{\prime}+\delta\rho_{E}^{\prime})}%
		{|\mathbf{q}-\mathbf{q}^{\prime}|}
		\mathrm{d}^3\mathbf{q} \, \mathrm{d}^3\mathbf{q}^{\prime}
\label{decomposeh2}
\end{equation}
where
\begin{eqnarray}
	K_{L^{2}} &:=& -\int \partial_{L^{2}} f_0^a \{g,E\} \{g,L^{2}\} \mathrm{d}\mathbf{\Gamma}\\
	K_{E} & := &  -\int \partial_E f_0^a \{g,E\}^{2} \mathrm{d}\mathbf{\Gamma}\\
	\delta\rho_{L^{2}} & := & - m \int \partial_{L^{2}} f_0^a \{g,L^{2}\} \mathrm{d}^3\mathbf{p}\\
	\delta\rho_{E} & := & - m \int \partial_E f_0^a \{g,E\} \mathrm{d}^3\mathbf{p}
\end{eqnarray}

For a general perturbation, it is difficult to say more about the sign of 
$H^{(2)}$. However, the system could receive any kind of perturbations.
In order to go further,  we have to make some assumptions about $g$.
We already know, from section~\ref{stabproblem}, that a radial function will
not lead to an instability, and from section~\ref{cas-dep}, that dependency on
$E$ and $L^2$ plays no part. To find a $g$ function that works, we thus have to
consider a non-radial perturbation. We can consider a perturbation that
is axisymmetric around the $z$ axis:
\begin{equation}
	g(\mathbf{\Gamma}) = g(E,L^{2},\theta,p_{\theta})
\label{perturb-form}
\end{equation}
With this hypothesis:
\begin{eqnarray}
	\{g, L^2 \} & = & 
	2 p_\theta \frac{\partial g}{\partial \theta}
		+ 2 p_\phi^2 \frac{\cos(\theta)}{\sin^3(\theta)}
			\frac{\partial g}{\partial p_\theta}
	\\
	\{ g, E \} & = & \frac{1}{2 m r^2} \{ g, L^2 \}
\end{eqnarray}

With $f_0^a = \varphi (E) \delta_{a}\left(L^{2}\right)$, we get
\begin{eqnarray}
	\partial_E f_0^a &=& \varphi'(E)\delta_a(L^2)
	\\
	\partial_{L^2} f_0^a &=& - \frac{1}{a^2} \varphi(E) \delta_a(L^2)
\end{eqnarray}
With the previous results, and using~(\ref{intdelta}), we can calculate
explicitly the four terms $K_{L^2}$, $K_E$, $\delta\rho_{L^2}$ and
$\delta\rho_{E}$ in a power series of $a$ for $a \rightarrow 0$.
If we consider only the first term in $a$, which corresponds to the term of
lower power in $p_\theta$ and $p_\phi$, a long but straightforward calculation
eventually leads to the following results:
\begin{eqnarray}
	K_{L^{2}} \! &\!=\!& \! \frac{1}{m} \int \frac{1}{r^{4}}
		\varphi(E)
		\left. \left( \frac{\partial g}{\partial\theta} \right)^{2} \right\vert_{L^{2}=0}
		\mathrm{d}p_{r}\,\mathrm{d}^3\mathbf{q}
	\label{KL2} \\
	K_{E} \! &\!=\!& \! - \frac{a^{2}}{m^2} \int \frac{1}{2r^{6}}
		\varphi^{\prime}(E)
		\left. \left( \frac{\partial g}{\partial\theta} \right)^{2}\right\vert_{L^{2}=0}
		\mathrm{d}p_{r} \, \mathrm{d}^3\mathbf{q}
	\label{KE} \\
	\delta\rho_{L^{2}} \! &\!=\!& \! \frac{m}{r^{2}}
		\int \varphi(E)
		\! \left. \left( \frac{\partial^{2}g}{\partial\theta\partial p_{\theta}}
			+ \frac{\cos(\theta)}{\sin(\theta)} 
				\frac{\partial g}{\partial p_{\theta}}
		\right)  \right\vert_{L^{2}=0} \!\! \mathrm{d}p_{r}
	\label{KRHOL2} \\
	\delta\rho_{E} \! &\!=\!& \! - \frac{m a^{2}}{2r^{4}} \! \int \! \varphi^{\prime}(E)
	\!\! \left. \left( \!
		\frac{\partial^{2}g}{\partial\theta \partial p_{\theta}}
		\! + \! \frac{\cos(\theta)}{\sin(\theta)} \!
			\frac{\partial g}{\partial p_{\theta}}
	\! \right) \right\vert_{L^{2} = 0}
	\!\!\! \mathrm{d}p_{r}
	\label{KRHOE}
\end{eqnarray}

As one can see $K_{E}$ and $\delta\rho_{E}$ are of order $a^{2}$ for
$a \rightarrow 0$, whereas $K_{L^{2}}$ and $\delta\rho_{L^{2}}$ do not depend on
$a$ in the same regime. Therefore, for near radial orbit systems one can
neglect $K_{E}$ in front of $K_{L^{2}}$ and $\delta\rho_{E}$ in front of
$\delta\rho_{L^{2}}$.

The second-order energy variation of perturbed near radial orbit systems is
then
\begin{eqnarray}
	H^{(2)}[f_0^a] & = &
	\frac{1}{m}  \int \frac{ \varphi(E) }{r^{4}}
		\left. \left(  \frac{\partial g}{\partial\theta} \right)^{2} \right\vert _{L^{2}=0}
		\mathrm{d}p_{r}\,\mathrm{d}^3\mathbf{q}
		\nonumber \\
	& - & G \int\!\!\int \frac{\delta\rho_{L^{2}} \delta\rho_{L^{2}}^{\prime}}%
		{|\mathbf{q} - \mathbf{q}^{\prime}|}
		\mathrm{d} \mathbf{q} \, \mathrm{d} \mathbf{q}^{\prime}
\end{eqnarray}

The first term is clearly positive as an integral of a positive function, while
the second is clearly negative owing to the negativeness of the Laplacian
operator: introducing
\begin{equation}
	\mu(\mathbf{q}) :=
	- \int \frac{\delta\rho_{L^{2}}^{\prime}}{|\mathbf{q}-\mathbf{q}^{\prime}|}
	\mathrm{d}^3\mathbf{q}^{\prime}
\end{equation}
one has $\Delta \mu = 4\pi\delta\rho_{L^{2}}$, hence
\begin{eqnarray*}
	- \int\!\!\int
	\frac{\delta\rho_{L^{2}}\delta\rho_{L^{2}}^{\prime}} {|\mathbf{q}-\mathbf{q}^{\prime}|}
	\mathrm{d}^3\mathbf{q} \, \mathrm{d}^3\mathbf{q}^{\prime}
	& = & \frac{1}{4\pi} \int\mu\Delta\mu \mathrm{d} \mathbf{q} \\
	& = & -\frac{1}{4\pi} \int(\nabla\mu)^{2} \mathrm{d}^3\mathbf{q}
	< 0
\end{eqnarray*}

The sign of $H^{(2)}$ is thus unclear, unless we make another assumption about
$g$. To be able to say more, we can consider a generating
function verifying
\begin{equation}
	\forall E,\ \theta \ :
	\quad \left. \frac{\partial g}{\partial\theta}\right\vert_{L^{2}=0}
	= 0
	\quad \text{while} \quad
	\left. \frac{\partial g}{\partial p_{\theta}} \right\vert_{L^{2}=0}
	\neq 0
\label{cond-gen}
\end{equation}

For such perturbations one can easily check that $K_{L^{2}}=0$ and
$\delta \rho_{L^{2}} \neq 0$, therefore
\begin{equation}
	H^{(2)}[f_0^a]
	= - G \int\!\!\int \frac{\delta\rho_{L^{2}}\delta\rho_{L^{2}}^{\prime}}%
		{|\mathbf{q}-\mathbf{q}^{\prime}|}
	\mathrm{d}^3\mathbf{q} \, \mathrm{d}^3\mathbf{q}^{\prime}
	< 0
\end{equation}

To summarize, in this section, we have obtained:

\begin{result}
Let $f_0^a$ be a distribution function of nearly-radial orbits, such that
$$f_{0}^{a} = \varphi\left( E\right)
\frac{1}{\pi a^{2}} \exp\left(-\frac{L^{2}}{a^{2}}\right).$$
Let $g$ be a generator of a perturbation, of the form
$g(\mathbf{\Gamma}) = g(E,L^{2},\theta,p_{\theta})$, with 
$\left. \frac{\partial g}{\partial\theta}\right\vert_{L^{2}=0} = 0$
while
$\left. \frac{\partial g}{\partial p_{\theta}} \right\vert_{L^{2}=0} \neq 0$.

Then for sufficiently small values of $a$, the energy variation $H^{(2)}[f_0^a]$
caused by $g$ is negative.
\end{result}

%---------------------------------------------------------------------

\subsection{Density variation}

It is interesting to analyse the density variation associated to the
generator described in the previous section.
The symplectic formulation of the problem allows us to
write the perturbation of the distribution function in terms of the generating
function $g$: this is equation~(\ref{symp-pert}).
From this relation one can obtain the density
\begin{eqnarray*}
	\rho(\mathbf{q}) &=& \rho_{0} + \rho^{(1)} = \int m f \mathrm{d} \mathbf{p}
	\\
	&=& \int m f_0^a \mathrm{d}^3\mathbf{p}
		- \int m \{g,f_0^a\} \mathrm{d}^3\mathbf{p}
	\\
	&=& \int m f_0^a \mathrm{d}^3\mathbf{p}
		+ \delta \rho_E + \delta \rho_{L^2}
\end{eqnarray*}

For sufficiently small values of $a$, $\delta\rho_{E}$ is negligible in front
of $\delta\rho_{L^{2}}$, as we have seen in~(\ref{KRHOE}),
hence $\delta^{(1)}\rho = \delta\rho_{L^{2}}$.
Using~(\ref{KRHOL2}), it can be checked that the first-order variation
of total mass $\delta^{(1)} m$ associated to the perturbation is vanishing.
\begin{eqnarray*}
	\!\! \delta^{(1)}m \!\!
	&\!\! = \!\!& \int \delta^{(1)}\rho \mathrm{d}^3\mathbf{q}
	= m \int \delta \rho_{L^{2}} \, r^{2} \sin(\theta)
		\mathrm{d}r \, \mathrm{d}\theta \, \mathrm{d}\phi\\
	&\!\! = \!\!& m \int \varphi(E)
		\left. \left( \sin(\theta)
			\frac{\partial^{2}g}{\partial\theta\partial p_{\theta}}
			+ \cos(\theta)
			\frac{\partial g}{\partial p_{\theta}}
		\right) \right\vert_{L^{2}=0} \!\!\!\!\!
		\mathrm{d}p_{r} \, \mathrm{d}r \, \mathrm{d}\theta \, \mathrm{d}\phi\\
	&\!\! = \!\!& m \int \varphi(E)
		\left( \int_{0}^{\pi} \frac{\partial}{\partial\theta}
			\left( \sin(\theta)\frac{\partial g}{\partial p_{\theta}}\right)
			\mathrm{d}\theta
		\right) \mathrm{d}p_{r} \, \mathrm{d}r \, \mathrm{d}\phi \\
	&\!\! = \!\!& 0
\end{eqnarray*}
Without more hypotheses than~(\ref{cond-gen}) on the perturbation
generating function $g$, and for sufficiently small values of $a$, the first
order induced variations of density are
\[
	\delta^{(1)} \rho = \delta\rho_{L^{2}}
	= \frac{m}{r^{2}} \int\varphi(E)
	\left.  \left( \frac{\partial^{2}g}{\partial\theta\partial p_{\theta}}
		+ \frac{\cos(\theta)}{\sin(\theta)}
			\frac{\partial g}{\partial p_{\theta}}
	\right) \right\vert_{L^{2}=0} \mathrm{d}p_{r}
\]

In order to obtain some physical characteristics of the instability,
we have to make yet another assumption about $g$.
To find a function that verifies condition~(\ref{cond-gen}), given the form of
$g$ given in~(\ref{perturb-form}),
we can suppose for example that $g$ is separated in the $\theta$ variable, i.e.
one can find two functions $A$ and $B$ such that
\begin{equation}
	g(E,L^{2},\theta,p_{\theta}) = B(E,L^{2},p_{\theta}) A(\theta)
\label{variable-separation}
\end{equation}

Under this assumption, criterion~(\ref{cond-gen}) becomes
\begin{equation}
	\forall E,\ \theta\ :
	\quad
	A'(\theta) B(E,0,0)
	= 0
	\quad \text{while} \quad
	A(\theta) \frac{\partial B}{\partial p_\theta} (E,0,0)
	\neq 0
\end{equation}
It is very easy to find a $B$ that verifies this condition.
A direct calculation then gives
\[
	\delta^{(1)}\rho = m \frac{D \left(\theta\right)}{r^{2}}
		\int \varphi(E)|_{L^{2}=0} \frac{\partial B}{\partial p_\theta} (E,0,0) \mathrm{d}p_{r}
\]
where
\[
	D\left(\theta\right)
		= A^{\prime}(\theta) + \frac{\cos(\theta)}{\sin(\theta)}A(\theta)
\]

If $D\left(\theta\right)$ is not constant,
which corresponds to a wide class of $A$%
\,\footnote{The equation $D\left( \theta\right) \neq k$ can be easily solved
and gives
\[
	A(\theta)\neq\frac{\lambda - k\cos(\theta)}{\sin(\theta)}
\]
where $k$ and $\lambda$ are $\theta$-free constants.},
then $\delta^{(1)}\rho$ does depend on
$\theta$ and the spherical symmetry of the equilibrium state is broken.%
\,\footnote{The fact that the resulting perturbation depends only on $\theta$,
$r$ and $E$, and thus is axisymmetric around the $z$ axis, is of course a
consequence of our choice of the form~(\ref{perturb-form}) for the
generating function.}

We have reached our goal: we have found a class of perturbations $g$ which leads
to a negative energy variation, and which creates a density variation that is
\emph{not} spherically symmetric. As per section~\ref{section-method}, this
means that with the help of dissipation, the system is unstable against this
perturbation: hence a favoured direction will appear in the
system, which was initially spherical.

\begin{result}
Consider a self-gravitating system, described by the CBP system and represented
by a distribution function $f_0(E,L^2)$, that is spherically symmetric and with 
nearly radial orbits. Assume this system can dissipate energy.

Then there exists perturbations, generated by a function $g$, against which the
system is unstable, and that cause it to lose its spherical
symmetry.
\end{result}

%---------------------------------------------------------------------

\section{Conclusion}
In this paper we have shown two important points:
self-gravitating dynamical systems described by the Collisionless
Boltzmann--Poisson equations are candidates for Dissipation-Induced Instability
when they are more and more radially anisotropic;
and this mechanism generically introduces a favoured direction in the spatial
part of the system's phase space.
In comparison with previous tedious normal modes techniques used in this context,
the detail of the first point gives a simple proof of radial orbit
instability based on energetics arguments.
Dissipation, which is needed in our proof, is also implicitly required 
in the classical intuitive understanding of this instability presented in
\citet{Palmer} (section 7.3.1). It is a fact that in a {\sl pure} radial system
--- which is the most unstable --- orbits, which are frozen in a fixed
direction, cannot precess or librate as it is required for the trapping
resonance invoked by Palmer. Hence, if two radial orbits actually attain
a lower energy state by approaching each other, this mechanism actually needs
a way to dissipate excessive energy.
Finally, a point about time-scales should be stressed:
it is well known that radial orbit instability is effective on a few crossing
times, therefore if dissipation appears to be the cornerstone of radial orbit
instability, it is clear that it could not act alone.
Non-linear and non-local aspects of the gravitational potential clearly
amplifies and completes the dissipation-triggered work.

\label{section-conclusion}

\end{document}